\documentclass[%
 reprint,
 amsmath,amssymb,amsthm,
 nofootinbib,
 aps,
]{revtex4-1}

\usepackage{dcolumn}% Align table columns on decimal point
\usepackage{bm}% bold math
\usepackage[mathlines]{lineno}
\usepackage{graphicx}
\usepackage{csquotes}
\usepackage[final]{changes}

\begin{document}
\preprint{APS/123-QED}

\title{Simultaneous Quantization and Reduction of Constrained Systems}
%\title{Quantizing constrained Systems Using the Extended Stationary Action Principle}

\author{Jianhao M. Yang}
\email[]{jianhao.yang@alumni.utoronto.ca}
\affiliation{Qualcomm, San Diego, CA 92321, USA}

\date{\today}		

\begin{abstract}
We present a novel framework for quantizing constrained quantum systems in which the processes of quantization and constraint enforcement are performed simultaneously. The approach is based on an extension of the stationary action principle, incorporating an information-theoretic term arising from vacuum fluctuations. Constraints are included directly in the Lagrangian via Lagrange multipliers, allowing the subsequent variational procedure to yield the quantum dynamics without ambiguity regarding the order of quantization and reduction. We demonstrate the method through two examples: (i) a one-dimensional system with vanishing local momentum, where the simultaneous approach produces the time-independent Schrödinger equation while conventional reduced and Dirac quantization yield only trivial states, and (ii) a bipartite system with global translational invariance, where all three methods agree. These results show that the proposed framework generalizes standard quantization schemes and provides a consistent treatment of systems with constraints that cannot be expressed as linear operators acting on the wave function. In addition to a unified variational principle for constrained quantum systems, the formalism also offers an information-theoretic perspective on quantum effects arising from vacuum fluctuations.
%It is well known that Dirac quantization and reduced quantization are inequivalent for quantizing a constrained system, owning to the ambiguity in whether one should quantize before imposing constraints or vice versa. To eliminate this ambiguity, we propose a novel approach that implements quantization and constraint simultaneously. This method is based on a recently developed mathematical framework that quantizes classical systems via the extended stationary action principle. We apply the new approach to quantize two simple one-dimensional constrained systems. For a bipartite system with global translational invariance, the quantization results are consistent with those of reduced quantization or Dirac quantization. However, for an ensemble with vanishing local momentum, where the constraint is not linear with respect to the wave function, reduced quantization or Dirac quantization yields only a trivial solution with constant probability density. In contrast, our method additionally provides the time-independent Schrödinger equation, along with the trivial solution. We argue that this simultaneous quantization–constraint framework provides a more general and robust method for quantizing constrained systems.
\end{abstract}
%In other words, Bell inseparability is an informational consequence due to local vacuum fluctuations. 
\maketitle
%\onecolumngrid
%========================================
%========================================
\section{Introduction}
Quantization of constrained Hamiltonian systems is an important subject in modern physics. In gauge theories, for example, gauge symmetry appears as the invariance of the classical Lagrangian under local transformations, which induces constraints on the system’s physical variables\cite{Dirac, Isham, Hanson, McMullan, Dolan}. Invariance signals the presence of unphysical degrees of freedom in the configuration space, and their elimination reduces the phase space. The central issue is the ordering of reduction and quantization. The \textit{reduced quantization}~\cite{Isham} solves the constraints first at the classical level, then quantizes the reduced system. The \textit{Dirac quantization}~\cite{Dirac}, on the other hand, quantizes the unconstrained system first, then the constraints are solved at the quantum level. These two procedures are illustrated schematically in Fig. 1. Whether quantization and reduction commute is a subtle and highly nontrivial question, and in general they do not. Numerous studies have shown explicit cases where the two approaches yield inequivalent quantum theories\cite{Ashtekar, Loll, Romano, Schleich, Kunstatter}. In some cases, the Dirac quantization gives the energy spectra that match the experimental observations \cite{Romano}. Quantizing the constraints themselves can introduce quantum effects that are absent in the classically reduced theory. However, no rigorous proof exists that constraint quantization is universally self-consistent.

Consider how a constraint, denoted $\phi$, is implemented in Dirac quantization \cite{Dirac}. The procedure begins by canonically quantizing the unconstrained system. The state is represented by a wave function $\Psi$ defined on the full configuration space, whose dynamics obey the Schrödinger equation. The constraint is then promoted to an operator $\hat{\phi}$, and the physical subspace is defined by imposing
\begin{equation}
\label{qconst}
    \hat{\phi}\Psi =0.
\end{equation}
Eq.\eqref{qconst} is a strong condition. It assumes that $\phi$ admits a consistent operator representation and further requires $\hat{\phi}$ to act linearly on the wave function. The latter ensures that any superposition of solutions to \eqref{qconst} is itself a solution. In practice, Eq. \eqref{qconst} functions as an additional postulate of Dirac quantization rather than a consequence of the standard postulates of quantum mechanics. This raises the question of whether there is a more natural formulation, one that avoids introducing such an ad hoc condition and resolves the ambiguity in the ordering of quantization and reduction.

\begin{figure*}
\begin{center}
\includegraphics[scale=0.5]{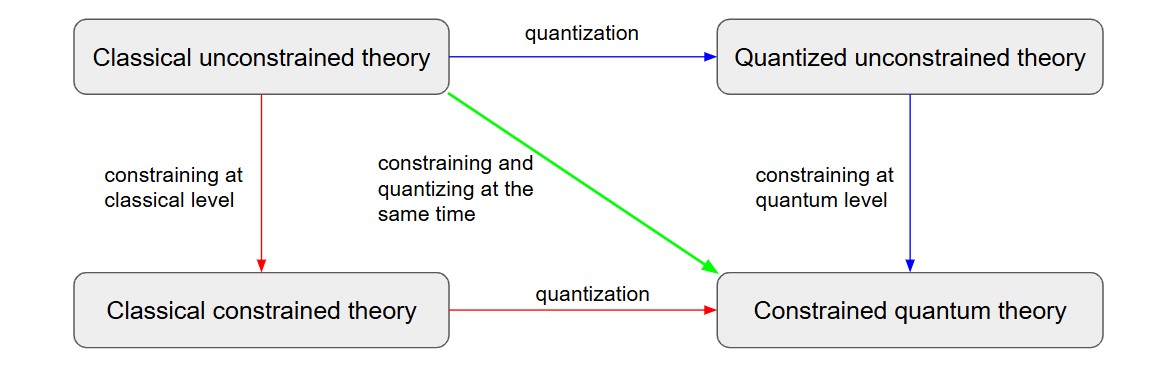}
\caption{Three procedures to derive quantum theory for a constrained system.}
\end{center}
\label{fig:constrained}
\end{figure*}

The aim of this work is to propose an alternative framework for the quantization of constrained systems. The key idea is to implement quantization and reduction simultaneously, thereby eliminating the ambiguity of whether constraints should be imposed before or after quantization. In this formulation, there is no need to quantize the constraints separately, and consequently, the condition in Eq.~\eqref{qconst} is not required. The approach is grounded in the extended stationary action principle for quantum mechanics~\cite{Yang2023}, which generalizes the classical principle of stationary action by introducing an information metric $I_f$ that accounts for additional observable information arising from random vacuum fluctuations. The dynamical laws follow from extremizing the combined action of the classical trajectory and the vacuum fluctuations. When the principle is applied recursively over infinitesimal and finite time intervals, the uncertainty relation and the Schrödinger equation are recovered, respectively. This framework has been successfully applied to derive the nonrelativistic Schrödinger equation~\cite{Yang2023}, electron spin~\cite{Yang2025}, scalar field theory~\cite{Yang2024}, and fermionic field theory~\cite{Yang2025_2}. Conceptually, it attributes quantumness to the information metric $I_f$; Mathematically, it provides a variational framework in which quantization is derived from a Lagrangian formulation rather than imposed via canonical commutation relations. This mathematical framework offers a new possibility that physical constraints can be incorporated naturally through Lagrange multipliers, allowing quantization and constraining to be carried out in a single step, as illustrated schematically in Fig.~1.

We expect this framework to more accurately capture the quantum behavior of systems with certain classes of constraints. To illustrate the new framework, we first consider a one-dimensional ensemble with the constraint of vanishing local momentum. In this case, reduced quantization and Dirac quantization yield only a trivial state with constant probability density, whereas the new approach produces, in addition, energy eigenstates governed by the time-independent Schrödinger equation. We then study a one-dimensional bipartite system subject to global translational invariance, where all three quantization methods give the same theory.The key distinction is that in the first example the constraint cannot be cast in the form of Eq.~\eqref{qconst}, while in the second it can, once quantization is performed within the new framework.

For clarity, we restrict our analysis to simple one-dimensional constrained systems. However, the framework is expected to generalize to more complex cases. We note that the first example of an ensemble with vanishing local momentum was previously studied by Hall~\cite{Hall}, who showed that Dirac quantization fails to implement such constraints, leading to a superselection of energy eigenstates. However,~\cite{Hall} still employs the two-step procedure of quantizing followed by constraining. In contrast, the present work integrates quantization and constraining into a single step, yielding a framework that is more general and capable of treating constrained systems beyond the reach of reduced or Dirac quantization.

The remainder of this paper is organized as follows. Section~\ref{LIP} reviews the extended stationary action principle, its assumptions, and the five-step derivation of quantum mechanics for an unconstrained system. Section~\ref{sec:ex1} applies the framework to a one-dimensional system with vanishing local momentum, yielding a non-trivial solution absent in reduced and Dirac quantization. Section~\ref{sec:constrainedBP} extends the analysis to a bipartite system with global translational invariance, where the new simultaneous approach reproduces the results of the conventional methods. Section~\ref{sec:discussion} discusses the broader applicability of the simultaneous method, provides an information-theoretic interpretation of the Bohm quantum potential, and presents our conclusions.

%\section{Dirac Quantization and Reduced Quantization}

\section{Quantization Using the Extended Stationary Action Principle}
\label{LIP}
\subsection{The Extended Stationary Action Principle}
The theoretical framework in this paper is developed based on the extended least action principle proposed in~\cite{Yang2023}. In essence, the principle of least action in classical mechanics is generalized to derive quantum formulations by incorporating the following two assumptions:
\begin{displayquote}
\emph{Assumption 1 -- A quantum system experiences vacuum fluctuations constantly. The fluctuations are local and completely random.}
\end{displayquote}
\begin{displayquote}
\emph{Assumption 2 -- There is a lower limit to the amount of action that a physical system needs to exhibit in order to be observable. This basic discrete unit of action effort is given by $\hbar/2$ where $\hbar$ is the Planck constant.}
\end{displayquote}
The conceptual justifications of these two assumptions have been extensively discussed in \cite{Yang2023}. Here we just briefly review these justifications. The first assumption aligns with the standard view that vacuum fluctuations underlie the intrinsic randomness of quantum dynamics. Although we do not know the physical details of the vacuum fluctuation, the crucial assumption here is the locality of the vacuum fluctuation. This implies that for a composite system, the fluctuation of each subsystem is independent of each other. Assumption 2 provides us with a new mechanism to calculate the additional action due to vacuum fluctuations. Although the microscopic details of the fluctuations remain unknown, the vacuum fluctuations manifest themselves via a discrete action unit determined by the Planck constant as an observable information unit. If an information metric is defined to measure the observable information manifested by vacuum fluctuations, multiplying this metric by the Planck constant yields the action associated with vacuum fluctuations. Then, the challenge of calculating the additional action due to the vacuum fluctuation is converted to defining an appropriate new information metric $I_f$. The task of defining an appropriate information metrics is less challenging since there are information-theoretic tools available. The most commonly used information metrics that extract observable information are defined by relative entropy. The concrete form of $I_f$ will be defined later as a functional of the Kullback-Leibler divergence $D_{KL}$, $I_f:=f(D_{KL})$, where $D_{KL}$ measures the information distances of different probability distributions caused by vacuum fluctuations. Thus, the total action is
\begin{equation}
\label{totalAction}
    A_t = A_c + \frac{\hbar}{2}I_f,
\end{equation}
where $A_c$ is the classical action. Quantum theory can be derived~\cite{Yang2023, Yang2024} through a variation approach to extremize such a functional quantity, $\delta A_t=0$. In the classical limit $\hbar \to 0$, $A_t \to A_c$, and extremization recovers classical dynamics. For $\hbar \neq 0$, the contribution of $I_f$ must be retained, and this information metric becomes the source of quantum behavior. These considerations may be summarized as\footnote{Over its development \cite{Yang2023, Yang2024, Yang2025, Yang2025_2}, the principle has been referred to by various names, including the principle of least observability and the extended principle of least action. These changes reflect a progressively refined understanding of its content.}
\begin{displayquote}
\emph{\textbf{Extended Stationary Action Principle} -- The dynamics of a quantum system extremize the action functional defined in (\ref{totalAction}).}
\end{displayquote}

Within this framework, the quantization of a classical system proceeds through the following five steps:

\begin{itemize}
    \item \textbf{Step I} Write down the classical Lagrangian for the system to be quantized. %For field theory, write down the classical Lagrangian density as that in the standard canonical quantization.
    \item \textbf{Step II} Apply the extended stationary action principle for an infinitesimal short time step. Define the relative entropy between the transition probability induced by vacuum fluctuations and a uniform reference distribution. The variation of the total action \eqref{totalAction} yields the transition probability density of the vacuum fluctuation. From the probability density, the variance of vacuum fluctuations can be computed.
    \item \textbf{Step III} Perform a canonical transformation to obtain the Hamilton-Jacobi equation with the generating function $S$. Introduce a configuration space for an ensemble of systems with probability density $\rho$. Compute the classical action $A_c$ for the ensemble. 
    \item \textbf{Step IV} Apply the extended stationary action principle again for a period of time to extract the dynamic equation for the ensemble. The relative entropy between the distributions with and without fluctuations for the period of time $t\in\{0, T\}$ contributes an additional term $I_f$ in the total action $A_t$.
    \item \textbf{Step V} Carry out the variation over $A_t$ with respect to $\rho$ and $S$ resulting in two differential equations for the dynamics of $\rho$ and $S$. Defining the wave function $\Psi = \sqrt{\rho}, e^{iS/\hbar}$, these equations combine to obtain the Schrödinger equation.
\end{itemize}
Steps~I and III remain within the classical domain, while quantization occurs in Steps II and IV by applying the same extended stationary action principle. Step V is mostly mathematical calculation. This framework has been successfully applied to derive a series of quantum theories, including non-relativistic quantum mechanics~\cite{Yang2023}, electron spin theory~\cite{Yang2025}, scalar and fermionic field theories~\cite{Yang2024, Yang2025_2}.

\subsection{Quantizing a One-dimensional System}
\label{appendix:qm}
In this subsection, we review the quantization of a nonrelativistic system. Although the theory has been developed previously~\cite{Yang2023}, our purpose here is to explicitly illustrate the five-step procedure outlined in the preceding subsection, thereby laying the groundwork for the later discussion of constrained systems. For clarity, we consider the simplest case: the quantization of a one-dimensional system.

First, the Lagrangian of a single one-dimensional classical system is simply
\begin{equation}
\label{LS}
    L = \frac{1}{2}m\dot{x}^2 - V(x).
\end{equation}
The momentum $p=m\dot{x}$ and spatial coordinates $x$ form a pair of canonical coordinates. The Hamiltonian is given by $H(x, p) = p\dot{x}-L=p^2/2m + V(x)$. 

In Step II, we consider the dynamics of a system with an infinitesimal time internal $\Delta t$ due to vacuum fluctuation. Define the probability that the system will transition from a spatial position ${x}$ to another position ${x}+{w}$ as $\varrho({x}+{w}|{x})d{w}$, where ${w}=\Delta {x}$ is the displacement due to fluctuations. The expectation value of the classical action is $A_c=\int \varrho({x}+{w}|{x})Ld{w}dt$. For an infinitesimal time internal $\Delta t$, one can approximate the velocity ${v}={w}/\Delta t$. This gives 
\begin{equation}
\label{action1}
    A_c=\int^{+\infty}_{-\infty}\varrho(\frac{m}{2\Delta t}w^2 + V\Delta t) d{w}.
\end{equation}
The second term $V\Delta t$ can be ignored when $\Delta t\to 0$. The information metrics $I_f$ is defined as the Kullback–Leibler divergence, to measure the information distance between $\varrho({x}+{w}|{x})$ and a uniform prior probability distribution $\sigma$ that reflects that the vacuum fluctuations are completely random with maximal ignorance~\cite{Jaynes}, 
\begin{align*}
    I_f  =: D_{KL}(\varrho({x}+{w}|{x}) || \sigma) 
    = \int \varrho \ln(\varrho/\sigma)d{w}.
\end{align*}
Inserting both $A_c$ and $I_f$ into (\ref{totalAction}), we have
\begin{equation}
    A_t = \int\varrho(\frac{m}{2\Delta t}w^2 +\frac{\hbar}{2} \ln\frac{\rho}{\sigma}) d{w}.
\end{equation}
Performing the variation procedure with respect to $\varrho$, one obtains
\begin{equation}
\label{transP}
    \varrho({x}+{w}|{x}) = \frac{1}{Z}e^{-\frac{m}{\hbar\Delta t}w^2},
\end{equation}
where $Z$ is a normalization factor. Equation (\ref{transP}) shows that the transition probability density is a Gaussian distribution. The variance is
\begin{equation}
\label{variance}
    \langle w^2\rangle = \frac{\hbar\Delta t}{2m}.
\end{equation}
Recalling that $w/\Delta t = v$ is the approximation of velocity due to the vacuum fluctuations, one can deduce
\begin{equation}
\label{exactUR}
    \langle\Delta x\Delta p\rangle = \hbar/2.
\end{equation}
Applying the Cauchy–Schwarz inequality gives the uncertainty relation,
\begin{equation}
    \langle\Delta x\rangle\langle\Delta p\rangle \ge \hbar/2.
\end{equation}

In the third step, we apply the canonical transformation in classical mechanics by introducing a pair of generalized canonical coordinates $\{X, P\}$. Denote $K(X, P)$ and $L'(X,P)$ the Hamiltonian and the Lagrangian in the generalized coordinate system. They are related via $L'= P\dot{X}-K$. The details of the extended canonical transformation are presented in Appendix \ref{appendix:canonical}, with the key results as following
\begin{align}
    p &= \frac{\partial S}{\partial x},\\
    L' &= \frac{\partial S}{\partial t} +H.
\end{align}
where $S$ is a generating function. The action integral in the generalized canonical coordinates becomes
\begin{equation}
    \label{extActionA}
    A_c = \int^{t_B}_{t_A}dt L' = \int^{t_B}_{t_A}dt \{\frac{\partial S}{\partial t} + H(x, \frac{\partial S}{\partial x})\}.
\end{equation}

In order to describe the time evolution of the systems that are subject to uncertainty, we adopt\footnote{More precisely, we only adopt the framework from \cite{HallBook} for the classical ensembles in Step III. The principle for quantization in Step IV is fundamentally different from \cite{HallBook}. This will be discussed in detail at the end of this section.} the theoretical framework of ensemble on configuration space, developed by Hall and Reginatto~\cite{HallBook}. In this framework, the uncertainty regarding the position of the system is described by a probability density over the configuration space, $\rho(x, t)$.  The Lagrangian density for the ensemble is naturally defined as $\mathcal{L}=\rho L'$. Thus, the classical action of the ensemble is
\begin{equation}
    \label{extActionE}
    A_e = \int dxdt \mathcal{L} = \int dxdt \rho \{\frac{\partial S}{\partial t} + H(x, \frac{\partial S}{\partial x})\}.
\end{equation}
As shown by Hall and Reginatto~\cite{Hall:2001,Hall:2002}, the Hamilton–Jacobi and continuity equations can be obtained from the classical action $A_e$ by fixed-point variation with respect to $\rho$ and $S$, respectively. In particular, taking variation of $A_e$ with respect to $\rho$ yields the Hamilton–Jacobi equation,
\begin{equation}
    \label{HJE}
    \frac{\partial S}{\partial t }+ \frac{1}{2m}(\frac{\partial S}{\partial x})^2 + V = 0.
\end{equation}
Taking variation of $A_e$ with respect to $S$ gives the continuity equation 
\begin{equation}
\label{contEqq}
    \frac{\partial\rho }{\partial t }+ \frac{1}{m}\frac{\partial}{\partial x}(\rho \frac{\partial S}{\partial x}) = 0.
\end{equation}
These results suggest that $(\rho, S)$ can also be considered as a pair of generalized canonical coordinates as well for the ensemble. To confirm this observation, first, the Lagrangian density of the ensemble is
\begin{equation}
    \mathcal{L}[\rho, S] = \rho \{\frac{\partial S}{\partial t} + H(x, \frac{\partial S}{\partial x})\}.
\end{equation}
It follows that
\begin{equation}
    \rho = \frac{\delta \mathcal{L}}{\delta(\partial_t S)}
\end{equation}
Thus, $\rho$ can be considered as the generalized conjugate momentum for $S$. Second, the classical ensemble Hamiltonian is given by the following functional
\begin{equation}
    \label{eHimiltonian}
    H_e[\rho, S] = \int dx \rho K=\int dx \rho[\frac{1}{2m}(\frac{\partial S}{\partial x})^2+V].
\end{equation}
One can verify that equations \eqref{HJE} and \eqref{contEqq} are equivalent to the equations of motion based on $H_e$,
\begin{equation}
   \frac{\partial S}{\partial t}=-\frac{\delta H_e}{\delta \rho} , \textbf{ }\frac{\partial\rho}{\partial t}=\frac{\delta H_e}{\delta S}.
\end{equation}
These results justify the choice of $(\rho, S)$ as a pair of generalized canonical coordinates that together with the ensemble Hamiltonian $H_e$, generates the dynamic equations of the classical ensemble. Eq. \eqref{extActionE} is the starting point for the quantization process in the next step.

Step IV, to define the information metrics for the vacuum fluctuations $I_f$, we slice the time duration $t_A\to t_B$ into $N$ short time steps $t_0=t_A, \ldots, t_j, \ldots, t_{N-1}=t_B$, and each step is an infinitesimal period $\Delta t$. In an infinitesimal time period at time $t_j$, the particle not only moves according to the Hamilton-Jacobi equation but also experiences random fluctuations. Such an additional revelation of distinguishability due to the vacuum fluctuations on top of the classical trajectory is measured by the following definition,
\begin{align}
\label{DLDivergence}
    I_f &=: \sum_{j=0}^{N-1}\langle D_{KL}(\rho ({x}, t_j) || \rho ({x}+{w}, t_j))\rangle_w \\
    &=\sum_{j=0}^{N-1}\int d{w}\varrho({w})\int dx\rho ({x}, t_j)\ln \frac{\rho ({x}, t_j)}{\rho ({x}+{w}, t_j)}.
\end{align}
When $\Delta t\to 0$, and with the identity \eqref{variance}, $I_f$ turns out to be~\cite{Yang2023}
\begin{equation}
\label{FisherInfo}
    I_f = \int d{x}dt \frac{\hbar}{4m}\frac{1}{\rho}(\frac{\partial\rho}{\partial x})^2.
\end{equation}
Eq. (\ref{FisherInfo}) contains the term related to Fisher information~\cite{Frieden, Reginatto} for the probability density $\rho$ but encodes more physical significance than Fisher information. %It shows that $I_f$ is proportional to $\hbar$. This is not trivial because it avoids introducing additional arbitrary constants for the subsequent derivation of Schr\"{o}dinger equation. Defining $I_f$ using other generic forms of relative entropy such as R\'{e}nyi divergence or Tsallis divergence, one will obtain a generalized Schr\"{o}dinger equation. 
Inserting (\ref{extActionE}) and (\ref{FisherInfo}) into (\ref{totalAction}), we obtain the total action
\begin{equation}
    \label{totalAction2}
    A_t = \int dxdt \rho\{ \frac{\partial S}{\partial t} + \frac{1}{2m}(\frac{\partial S}{\partial x})^2+V+\frac{\hbar}{4m}(\frac{\partial \ln\rho}{\partial x})^2\}.
\end{equation}

In step V, we perform the variation procedure on $A_t$ with respect to $S$, which gives the continuity equation. On the other hand, performing the variation with respect to $\rho$ leads to the quantum Hamilton-Jacobi equation,
\begin{align}
\label{QHJ}
    \frac{\partial S}{\partial t} + \frac{1}{2m}(\frac{\partial S}{\partial x})^2 + V +Q  = 0,\\
    \label{BohmP}
    \text{where } Q=- \frac{\hbar^2}{2m}\frac{\nabla^2\sqrt{\rho}}{\sqrt{\rho}}.
\end{align}
$Q(\rho)$ is called the Bohm quantum potential~\cite{Bohm1952}. We will discuss its physical interpretation in further detail in Section \ref{sec:Bohm}. Defining a complex function $\Psi=\sqrt{\rho}e^{iS/\hbar}$, the continuity equation and the extended Hamilton-Jacobi equation (\ref{QHJ}) can be combined into a single differential equation,
\begin{equation}
    \label{SE}
    i\hbar\frac{\partial\Psi}{\partial t} = [-\frac{\hbar^2}{2m}\nabla^2 + V]\Psi,
\end{equation}
which is the Schr\"{o}dinger Equation.

Once the Schr\"{o}dinger equation for the wave functional is derived and the correct Hamiltonian operator is identified, one can return to the standard operator-based approach. For instance, operators for momentum and angular momentum can be defined, and the energy of the ground state or excited states can be calculated. %The important point here is that the Schr\"{o}dinger equation is derived from the first principle rather than through a postulate in standard canonical quantization.

In the canonical quantization procedure, the position and momentum variables are promoted to operators and the commutation relation $[\hat{x},\hat{p}]=i\hbar$ is imposed, with the Schrödinger equation subsequently postulated. Although this prescription is mathematically elegant, it also appears somewhat ad hoc, as the operator promotion step lacks a deeper justification. In contrast, in the present framework, quantization is implemented by introducing an additional term $I_f$ into the Lagrangian and applying the standard variational principle. The Schrödinger equation then emerges directly from first principles, without the seemingly mystical step of promoting classical variables to operators. 

Although the presented quantization process adopts the framework for classical ensembles on configuration space developed in \cite{HallBook}, the underlying principle for quantization is fundamentally different. First, \cite{HallBook} postulates an ``exact uncertainty principle", which is \eqref{exactUR}. In our approach, \eqref{exactUR} is derived from the same first principle in both Steps II and IV, the extended stationary action principle. Second, in \cite{HallBook} the Bohm potential is derived from the random fluctuation of momentum and the exact uncertainty principle. In our framework, the Bohm potential is a consequence of the introduction of relative entropy $I_f$ to measure the information revealed from random fluctuations. This offers an information-theoretic perspective for the foundation of quantum theory. %Third, even within the classical ensemble framework, the ensemble Hamiltonian $H_e$ is introduced as a conjecture in \cite{HallBook}. Its justification is confirmed by its ability to generate the correct equations of motion. In our formalism, on the other hand, $H_e$ is derived through an extended canonical transformation.

With the variational formulation presented here, the treatment of constraints is straightforward: they can be incorporated into the Lagrangian through Lagrange multipliers in Steps II and IV, and the variation procedure is carried out in a single calculation. This provides a natural and unified procedure in which quantization and constraint enforcement occur simultaneously, thereby resolving the long-standing ambiguity regarding the order of these two operations. In the next two sections, we explore this approach for specific constrained systems and compare the results with those obtained using reduced quantization and Dirac quantization.

\section{Ensemble with Vanishing Local Momentum}
\label{sec:ex1}
Consider a constraint for an ensemble of single particles such that the local momentum of an ensemble is constant. This can be expressed as
\begin{equation}
\label{const19}
    \phi: p(x) - p_c\approx 0.
\end{equation}
The symbol $\approx$ indicates weak equality. That is, it is only valid on the constrained surface. We can always select a reference frame such that $p_c=0$. Hence, we will choose $p_c=0$, implying that the local momentum of the ensemble vanishes. For clarity, we retain the parameter $p_c$ in the intermediate steps of the formulation but set $p_c=0$ in the final stage. On the constrained surface of the phase space, the Hamiltonian is
\begin{equation}
\label{H10}
    H = \frac{p_c^2}{2m} + V(x).
\end{equation}
To preserve dynamics of the constraint, the Poisson bracket must vanish. Thus,
\begin{equation}
\label{const20}
    \{\phi, H\} = \frac{\partial\phi}{\partial x}\frac{\partial H}{\partial p}-\frac{\partial\phi}{\partial p}\frac{\partial H}{\partial x} \approx -\frac{\partial V}{\partial x} \approx 0.
\end{equation}
This is a secondary constraint, denoted as $\theta$.

\subsection{Reduced Quantization and Dirac Quantization}
With reduced quantization, we first apply the constraint $\phi$ at the classical level, and obtain the Hamiltonian \eqref{H10}. The constraint $\theta$ implies that $V(x) = V_c$ is a constant. Next, we perform canonical quantization. The promotion of the momentum operator $p\to\hat{p}$ does not change \eqref{H10}. The reduced Hamiltonian becomes
\begin{equation}
    H_{red} = \frac{p_c^2}{2m}+V_c = E.
\end{equation}
The Schr\"{o}dinger equation is simply
\begin{equation}
    i\hbar\frac{\partial\Psi}{\partial t} = E\Psi.
\end{equation}

With Dirac quantization, we first promote $p\to\hat{p}, x\to\hat{x}$, impose the commutative relation $[\hat{x}, \hat{p}]=i\hbar$, and postulate the Schr\"{o}dinger equation
\begin{equation}
    i\hbar\frac{\partial\Psi}{\partial t} = (\frac{\hat{p}^2}{2m} + V) \Psi.
\end{equation}
The two constraints are quantized as
\begin{align}
\label{const21}
    &\hat{\phi}: (\hat{p} - p_c)\Psi = 0, \\
\label{const22}
    &\hat{\theta}: (\frac{\partial V}{\partial x})\Psi = 0.
\end{align}
From \eqref{const21}, we get $\hat{p}^2\Psi = p_c^2\Psi$. And from \eqref{const22}, we obtain $\nabla V=0$ thus $V(x)=V_c$. Substituting them into the Schr\"{o}dinger equation, one has
\begin{equation}
    i\hbar\frac{\partial\Psi}{\partial t} = (\frac{p_c^2}{2m} + V_c) \Psi = E\Psi.
\end{equation}
Both reduced quantization and Dirac quantization give the same outcomes. But the resulting Schr\"{o}dinger equation admits a trivial solution,
\begin{equation}
    \Psi =\psi(x)e^{-iEt/\hbar}.
\end{equation}
where $\psi(x)$ is a real function. It is clear that the particle is in an energy eigenstate. Now we set $p_c=0$. The constraint \eqref{const21} states that $\hat{p}\Psi = 0$, which implies that $\partial\psi/\partial x = 0$ and $\psi$ is just a constant. Subsequently, $\rho= |\Psi|^2$ is also a constant. We find that the reduced quantization and Dirac quantization result in a trivial solution of constant probability density.

\subsection{Quantization Based on the Extended Stationary Action Principle}
\label{sec:ESAP}
Referring to Section \ref{appendix:qm}, we will go through each step in the quantization process, but incorporate constraint \eqref{const19}. The Lagrangian is still the same as \eqref{LS}. Step II is also the same as in Section \ref{appendix:qm}. In Step III, after the canonical transformation, the constraint \eqref{const19} becomes
\begin{equation}
    \label{const23}
    \phi\to\Phi: \frac{\partial S}{\partial x} - p_c \approx 0.
\end{equation}
After introducing the ensemble probability density $\rho$, the constraint $\phi$ leads to
\begin{align}
    \label{const24}
    \phi\to\Phi_e:& \int dx \rho(\frac{\partial S}{\partial x} - p_c) \approx 0.
%    \theta \to \Theta_e:&\int dx \rho(\frac{\partial V}{\partial x}) \approx 0.
\end{align}
When $p_c=0$, the local momentum of the ensemble vanishes. One should expect the probability density of the ensemble to be independent of time. This implies another constraint, $\partial\rho/\partial t \approx 0$. That is,
\begin{equation}
\label{qconst24}
    \Theta_e: \int dx \rho\frac{\partial \rho}{\partial t} \approx 0.
\end{equation}
For a reason to be clear later, we will perform the consistency checking after we obtain the Hamiltonian.

The crucial difference comes in Step IV. Here, the constraints are added to the Lagrangian  
\begin{equation}
    \label{totalL}
    \begin{split}
        L =& L_c+\lambda_1\Phi_e + \lambda_2\Theta_e\\
        =&  \int dx\rho\{\frac{\partial S}{\partial t}+H+\lambda_1(\frac{\partial S}{\partial x}-p_c )+\lambda_2\frac{\partial\rho}{\partial t} \}.
    \end{split}  
\end{equation}
where $\lambda_1$ and $\lambda_2$ are Lagrange multipliers. Then, the total action, after including the relative entropy term $I_f$, becomes 
\begin{equation}
\begin{split}  
    A_t =& \int dxdt\rho \{\frac{\partial S}{\partial t} +H +\lambda_1(\frac{\partial S}{\partial x}-p_c )+\lambda_2\frac{\partial\rho}{\partial t} \\
    &+\frac{\hbar}{4m}(\frac{\partial \ln\rho}{\partial x})^2  \}.
\end{split}
\end{equation}
This allows us to perform the quantization and solve the constrained conditions at the same time. Specifically, we perform the variation of $A_t$ with respect to $S, \rho$, $\lambda_1$, and $\lambda_2$, and obtain the following equations, respectively.
\begin{align}
    &\frac{\partial\rho }{\partial t }+ \frac{1}{m}\frac{\partial}{\partial x}(\rho \frac{\partial S}{\partial x}) +\lambda_1 \frac{\partial\rho }{\partial x }= 0,\\
    &\frac{\partial S}{\partial t} + H+Q +\lambda_1(\frac{\partial S}{\partial x}-p_c ) +\lambda_2\frac{\partial\rho}{\partial t}= 0, \\
    &\int dxdt \rho(\frac{\partial S}{\partial x}-p_c) =0, \int dxdt \rho\frac{\partial \rho}{\partial t} =0.
\end{align}
Substituting \eqref{const23} into the first two equations, we get
\begin{align}
\label{ccontEq}
    &\frac{\partial\rho }{\partial t }+ \frac{\partial}{\partial x}\{\rho(\frac{p_c}{m}+\lambda_1)\} = 0, \\
    &\frac{\partial S}{\partial t} + \frac{p_c^2}{2m} +V+Q +\lambda_2\frac{\partial\rho }{\partial t } = 0.
\end{align}
Eq. \eqref{ccontEq} is the continuity equation for the probability density of the constrained system. It shows that the current density of the ensemble is $j=\rho(p_c/m + \lambda_1)$. Next we substitute $\partial\rho/\partial t=0$ to the above two equations,
\begin{align}
\label{twoSol}
    &(\frac{p_c}{m}+\lambda_1)\frac{\partial\rho}{\partial x} = 0, \\
    &\frac{\partial S}{\partial t} + \frac{p_c^2}{2m} +V+Q  = 0.
\end{align}
There are two solutions for \eqref{twoSol}. Either $\partial\rho/\partial x = 0$ or $\lambda_1 = -p_c/m$. The first solution will lead to the same trivial result as the reduced or Dirac quantization. We are more interested in the second solution, which also fixes the value of $\lambda_1$. 

Now we set $p_c=0$ and obtain $\lambda_1=0$, and summarize the equations obtained from the variation procedure as following.
\begin{align}
\label{HJE2}
    &\frac{\partial S}{\partial t} +V + Q  = 0, \\
\label{ccontEq2}
    &\frac{\partial\rho }{\partial t } =0, \frac{\partial S}{\partial x}=0, \frac{\partial \rho}{\partial x} \ne 0.
\end{align}

In Step V, we wish to derive the Schr\"{o}dinger equation correspondent to the equations in \eqref{HJE2} and \eqref{ccontEq2}. Taking the time derivative of equation \eqref{HJE2}, one has
\begin{equation}
    \frac{\partial^2S}{\partial t^2} =-\frac{\partial Q(\rho)}{\partial t} = -\frac{\partial Q}{\partial \rho}\frac{\partial \rho}{\partial t} = 0.
\end{equation}
This implies that $\partial S/\partial t$ is a constant. Denote it as $-E$. Then, from \eqref{HJE2}, we get
\begin{equation}
\label{StaticHJE}
    V+Q = -\frac{\partial S}{\partial t} = E.
\end{equation}
This gives $S=-Et+\varphi$ where $\varphi$ is a constant. Substituting $Q$ from \eqref{BohmP} into \eqref{StaticHJE} gives
\begin{equation}
\label{HJE4}
    - \frac{\hbar^2}{2m}\frac{\nabla^2\sqrt{\rho}}{\sqrt{\rho}} + V = E.
\end{equation}
This can be rearranged as
\begin{equation}
    (- \frac{\hbar^2}{2m}\nabla^2 + V)\sqrt{\rho} = E\sqrt{\rho}.
\end{equation}
Defining $\Psi =\sqrt{\rho}e^{iS/\hbar} = \sqrt{\rho}e^{-iEt/\hbar}$ where we ignore the constant $\varphi$, one can verify that
\begin{equation}
\label{TimeIndSE}
    (-\frac{\hbar^2}{2m}\nabla^2 +V)\Psi = E\Psi.
\end{equation}
This is the time-independent Schr\"{o}dinger equation for the one-dimensional ensemble with constraint \eqref{const23}. It is the same result as in \cite{Hall}. Examples of systems governed by this equation include the one-dimensional quantum well and the one-dimensional quantum harmonic oscillator. The constrained system is in the energy eigenstate governed by \eqref{TimeIndSE}.

Eq.\eqref{HJE2} is the quantum version of Hamilton-Jacobi equation for the constrained system, analogous to the classical version of $\partial S/\partial t + H=0$. From \eqref{HJE2}, we identify the Hamiltonian $H=V+Q$ and the Hamiltonian of the quantum ensemble as
\begin{equation}
\label{HC}
    H_e = \int dx \rho(V+Q).
\end{equation}
In Appendix \ref{appdx:constcheck}, we verify that Poisson brackets $\{\Phi_e, H_e\} \approx 0$ and $\{\Theta_e, H_e\} \approx 0$. Thus, both constraints \eqref{const24} and \eqref{qconst24} are consistent and do not induce secondary constraints.

The key distinction between this solution and the trivial solution of Dirac quantization lies in the spatial derivative of the probability density: here, $\partial\rho/\partial x \ne 0$, whereas in the trivial solution $\partial\rho/\partial x = 0$. Since $\partial\rho/\partial x \ne 0$, the condition $\hat{p}\Psi = 0$ does not hold. In Dirac quantization, the constraint \eqref{const19} is promoted in \eqref{const21} to an operator that acts linearly on the wave function, which implicitly enforces $\partial\rho/\partial x = 0$ and excludes the second solution of \eqref{twoSol}. Consequently, the time-independent Schrödinger equation \eqref{TimeIndSE} cannot be derived within the Dirac framework.

This example makes clear that the extended stationary action principle provides a more general framework for quantizing systems with constraints that cannot be expressed as linear operators acting on the wave function. To see this explicitly, the constraint $\partial S/\partial x \approx 0$ can be re-expressed in terms of $\Psi$ as
\begin{equation}
\hat{p}(\ln\Psi - \ln\Psi^*) = 0,
\end{equation}
which differs fundamentally from the Dirac quantization condition $\hat{p}\Psi = 0$.

\section{Bipartite System with Translational Invariance}
\label{sec:constrainedBP}
An important class of constraints in physical systems arises from gauge symmetries. For instance, a Lagrangian exhibiting global translational invariance implies that the total center-of-mass momentum must vanish~\cite{Hoehn2018}. In the case of a one-dimensional $N$-particle system,
\begin{equation}
    \sum_{i=1}^{N}p_i \approx 0,
\end{equation}
Thus, the momenta of the individual particles are no longer independent. The momentum of the \textit{i}th particle is given by
\begin{equation}
    \label{momentum}
    p_i = m_i(\dot{x}_i - v_c)
\end{equation}
where $v_c$ is the velocity of the center of mass of the system, defined as $v_c = \sum_im_i\dot{x}_i/\sum_im_i$. The Lagrangian and Hamiltonian are \cite{Hoehn2018}
\begin{align}
\label{BPL}
    L  &=\sum_{i=1}^{N}\frac{1}{2}m_ix_i^2 - \frac{1}{2}(\sum_{i=1}^Nm_i)v_c^2- V\\
    &= \sum_{i=1}^{N}\frac{p_i^2}{2m_i} - V(\{x_i-x_j\}^N_{i,j=1}),\\
    \label{BPH}
    H&=\sum_{i=1}^{N}p_i\dot{x}_i - L = \sum_{i=1}^{N}\frac{p_i^2}{2m_i} + V.
\end{align}
Here, we will study the most simple system with translational invariance, a one-dimensional bipartite system. The constraint is simply
\begin{equation}
\label{constrained}
    \phi: p_a+p_b \approx 0.
\end{equation}
The constraint $\phi$ defines a three-dimensional constrained surface in the original four-dimensional phase space. The Hamiltonian is simply
\begin{equation}
    \label{BPHC}
    H = \frac{p_a^2}{2m_a}+\frac{p_b^2}{2m_b}+V(x_a-x_b).
\end{equation}
We can verify that the constraint is preserved by the Poisson bracket $\{\phi, H\}=0$. Thus, no secondary constraints are induced to enforce the conservation of $\phi$ and this indicates that $\phi$ is a first-class constraint.

\subsection{Reduced Quantization and Dirac Quantization}
With the reduced quantization approach, one first solves the constraints at the classical level. Given $\phi$, we simply replace $p_a=-p_b$ in the Hamiltonian and get
\begin{equation}
    \label{BPHR}
    H_{red} = (\frac{1}{2m_a}+\frac{1}{2m_b})p_a^2+V(x_a-x_b).
\end{equation}
Then we apply the canonical quantization by promoting 
\begin{equation}
    p_a\to\hat{p}_a, x_a\to\hat{x}_a, x_b\to\hat{x}_b,
\end{equation}
and imposing the commutative relation
\begin{equation}
    [\hat{x}_a, \hat{p}_a] = i\hbar, [\hat{x}_b, \hat{p}_a]=0.
\end{equation}
The reduced Hamiltonian operator is
\begin{equation}
    \label{BPHRO}
    \hat{H}_{red} = (\frac{1}{2m_a}+\frac{1}{2m_b})\hat{p}_a^2+V(x_a-x_b),
\end{equation}
which is then used to postulate the Schr\"{o}dinger equation
\begin{equation}
\label{BPRS}
    i\hbar\frac{\partial\Psi}{\partial t}=\hat{H}_{red}\Psi.
\end{equation}

In the Dirac quantization approach, we first apply the canonical quantization by promoting
\begin{equation}
    p_a\to\hat{p}_a, p_b\to\hat{p}_b, x_a\to\hat{x}_a, x_b\to\hat{x}_b,
\end{equation}
and imposing the commutative relation,
\begin{equation}
    [\hat{x}_\alpha, \hat{p}_\beta] = i\hbar\delta_{\alpha\beta}, \{\alpha,\beta=a,b\}.
\end{equation}
The quantization of Hamiltonian is straightforward,
\begin{equation}
    \label{BPH2}
    \hat{H}_{D} = \frac{\hat{p}_a^2}{2m_a}+\frac{\hat{p}_b^2}{2m_b}+V(x_a-x_b).
\end{equation}
The Schr\"{o}dinger equation is
\begin{equation}
\label{DiracSE}
    i\hbar\frac{\partial\Psi}{\partial t}=\hat{H}_{D}\Psi.
\end{equation}
Next step is to quantize the constraint $\phi$ by promoting 
\begin{equation}
    \phi\to\hat{\phi}=\hat{p}_a+\hat{p}_b,
\end{equation} 
and imposing
\begin{equation}
\label{Qconstrained}
    \hat{\phi}\Psi = (\hat{p}_a+\hat{p}_b)\Psi= 0.
\end{equation}
Lastly, one solves the constraint at the quantum level. From \eqref{Qconstrained}, we have $\hat{p}_a\Psi=-\hat{p}_b\Psi$. Then
\begin{equation}
    \hat{p}_a^2\Psi = -\hat{p}_a\hat{p}_b\Psi=-\hat{p}_b\hat{p}_a\Psi=\hat{p}_b^2\Psi.
\end{equation}
Substituting this into the Schr\"{o}dinger equation, we find
\begin{equation}
    i\hbar\frac{\partial\Psi}{\partial t}=(\frac{\hat{p}_a^2}{2m_a}+\frac{\hat{p}_a^2}{2m_b}+V(x_a-x_b))\Psi,
\end{equation}
which is identical to \eqref{BPRS}. We conclude that for such a constrained system, both reduced quantization and Dirac quantization give the same results.

\subsection{Quantization Based on the Extended Stationary Action Principle}
To illustrate how the constraint is incorporated within the framework of Sec.~\ref{LIP}, Appendix~\ref{appx:bipartite} first presents the quantization of the bipartite system without the constraint \eqref{constrained}. This serves as a reference for the modifications introduced in this section. Readers are strongly recommended to go through the Appendix~\ref{appx:bipartite} before proceeding.

We now impose the constraint \eqref{constrained}. The objective of Step II is to calculate the transition probability density of the system as a result of the fluctuation of the vacuum. The Lagrangian for the constrained bipartite system is given in \eqref{BPL}. Thus, the total action is similar to \eqref{A11} in Appendix~\ref{appx:bipartite} for the bipartite system without constraints, except with an extra term containing the velocity of the center of mass,
\begin{equation}
\label{BPTP}
\begin{split}
    A_t=&\int\varrho\{\frac{m_a}{2\Delta t}w_a^2 + \frac{m_b}{2\Delta t}w_b^2 - \frac{1}{2}Mv_c^2\Delta t+\frac{\hbar}{2}\ln\frac{\varrho}{\sigma}\} dw_adw_b
\end{split}
\end{equation}
where $M=m_a+m_b$. This extra term can be ignored since $v_c$ should be a finite quantity and when $\Delta t\to 0$, the term vanishes. Thus, \eqref{BPTP} is reduced to be the same as \eqref{A11}. Consequently, the transition probability density and the variance of vacuum fluctuations are identical to \eqref{BPp} and \eqref{BPvariance>}, respectively.

Since the forms of the constraint Lagrangian and Hamiltonian given in \eqref{BPL} and \eqref{BPH} are identical to those of the unconstrained bipartite system in terms of $p_a$ and $p_b$ , the canonical transformation in Step III proceeds exactly as in Appendix~\ref{appx:bipartite}. After this transformation, the constraint \eqref{constrained} takes the form
\begin{equation}
    \label{constrained2}
   \phi\to\Phi: \frac{\partial S}{\partial x_a}+\frac{\partial S}{\partial x_b} \approx 0.
\end{equation}
When we introduce the ensemble of the bipartite system with probability density $\rho$, the constraint \eqref{constrained2} implies
\begin{equation}
    \label{constrained3}
   \phi\to\Phi_e: \int dx_adx_b \rho(\frac{\partial S}{\partial x_a}+\frac{\partial S}{\partial x_b} )\approx 0,
\end{equation}
The Hamiltonian is
\begin{align}
\label{HeBP}
    H_e &= -\int dx_adx_b \rho H, \text{ where} \\
    H &=\frac{1}{2m_a}(\frac{\partial S}{\partial x_a})^2 + \frac{1}{2m_b}(\frac{\partial S}{\partial x_b})^2 + V(x_a-x_b).
\end{align}
In Appendix \ref{appdx:constcheck}, we show that the constraint \eqref{constrained3} is consistent with the equation of motion on the constrained surface and does not induce any secondary constraint. However, since the Lagrangian and Hamiltonian depend only on the relative position $(x_a-x_b)$, we expect that the probability density also depends only on the relative position $(x_a-x_b)$. That is, on the constrained surface,
\begin{equation}
    \rho(x_a, x_b, t) = \rho (x_a-x_b, t).
\end{equation}
This implies another constraint
\begin{equation}
\label{const2}
    \Theta: \frac{\partial \rho}{\partial x_a}+\frac{\partial \rho}{\partial x_b} \approx 0.
\end{equation}
For the ensemble of bipartite system, this constraint is written as
\begin{equation}
\label{constrained4}
    \Theta_e: \int dx_adx_b\rho(\frac{\partial \rho}{\partial x_a}+\frac{\partial \rho}{\partial x_b}) \approx 0.
\end{equation}
An interesting property of this constraint is
\begin{equation}
\begin{split}
    \delta\Theta_e &= \int dx_adx_b\{(\frac{\partial \rho}{\partial x_a}+\frac{\partial \rho}{\partial x_b})\delta\rho -(\frac{\partial \rho}{\partial x_a}+\frac{\partial \rho}{\partial x_b})\delta\rho \\
    &= 0.
\end{split}
\end{equation}
This implies that
\begin{equation}
    \frac{\delta\Theta_e}{\delta\rho} = 0, \frac{\delta\Theta_e}{\delta S}= 0.
\end{equation}
Consequently,
\begin{equation}
    \{\Theta_e, \Phi_e\} = 0, \{\Theta_e, H_e\}= 0.
\end{equation}
This indicates that there is no secondary constraint induced by $\Theta_e$. Thus, we have identified the complete set of constraints $\Phi_e$ and $\Theta_e$.

For Step IV, note that without constraints, the total action is given by \eqref{totalAction}. But with the constraints identified in \eqref{constrained3} and \eqref{constrained4}, we modify the Lagrangian as 
\begin{equation}
    \begin{split}
        L =& L_c+\lambda_1\Phi_e+\lambda_2\Theta_e \\
        =&  \int dx_adx_b\rho\{\frac{\partial S}{\partial t}+H+\lambda_1(\frac{\partial S}{\partial x_a}+\frac{\partial S}{\partial x_b} )\\
        &+\lambda_2(\frac{\partial \rho}{\partial x_a}+\frac{\partial \rho}{\partial x_b}) \}.
    \end{split}  
\end{equation}
where $\lambda_1, \lambda_2$ are the Lagrange multipliers. Then, the total action after including the relative entropy term $I_f$ becomes (compared to \eqref{A17})
\begin{equation}
\label{totalAction4}
\begin{split}  
    A_t =& \int dx_adx_bdt\rho \{\frac{\partial S}{\partial t} +H \\
    &+\lambda_1(\frac{\partial S}{\partial x_a}+\frac{\partial S}{\partial x_b} )+\lambda_2(\frac{\partial \rho}{\partial x_a}+\frac{\partial \rho}{\partial x_b})\\
    &+\frac{\hbar}{4m_a}(\frac{\partial \ln\rho}{\partial x_a})^2 + \frac{\hbar}{4m_b}(\frac{\partial\ln\rho}{\partial x_b})^2 \}.
\end{split}
\end{equation}
Next, taking the variations of $A_t$ with respect to $\rho$ gives
\begin{equation}
\label{BPHJC}
    \frac{\partial S}{\partial t} + H +Q_a+Q_b+\lambda_1(\frac{\partial S}{\partial x_a}+\frac{\partial S}{\partial x_b}) = 0.
\end{equation}
Performing the variations of $A_t$ with respect to $\rho$ results in
\begin{equation}
\label{BPCC}
\begin{split}
    \frac{\partial \rho}{\partial t} + \frac{1}{m_a}\frac{\partial}{\partial x_a}(\rho\frac{\partial S}{\partial x_a})^2 + \frac{1}{m_b}\frac{\partial}{\partial x_b}(\rho\frac{\partial S}{\partial x_b})^2& \\
     + \lambda_2(\frac{\partial \rho}{\partial x_a}+\frac{\partial \rho}{\partial x_b})&= 0 .
\end{split}
\end{equation}
Lastly, taking the variations of $A_t$ with respect to $\lambda_1$ and $\lambda_2$, just recovers the constraints \eqref{constrained3} and \eqref{constrained4}, respectively. Substituting \eqref{constrained2} into \eqref{BPHJC} one obtains
\begin{equation}
\label{BPHJC2}
    \frac{\partial S}{\partial t} + H +Q_a+Q_b = 0.
\end{equation}
Substituting \eqref{const2} into \eqref{BPCC} gives
\begin{equation}
\label{BPCC2}
    \frac{\partial \rho}{\partial t} + \frac{1}{m_a}\frac{\partial}{\partial x_a}(\rho\frac{\partial S}{\partial x_a})^2 + \frac{1}{m_b}\frac{\partial}{\partial x_b}(\rho\frac{\partial S}{\partial x_b})^2=0.
\end{equation}
As usual, defining a complex function $\Psi(x_a,x_b,t)=\sqrt{\rho(x_a,x_b,t)}e^{iS(x_a,x_b,t)/\hbar}$, Eqs. \eqref{BPHJC2} and \eqref{BPCC2} can be combined into the Schr\"{o}dinger equation,
\begin{equation}
    \label{BPCSE}
    i\hbar\frac{\partial\Psi}{\partial t} = [-\frac{\hbar^2}{2m_a}\nabla_a^2 -\frac{\hbar^2}{2m_b}\nabla_b^2+ V]\Psi.
\end{equation}
In terms of $\Psi$, Eqs. \eqref{constrained2} and \eqref{const2} are equivalent to
\begin{equation}
    \frac{\partial \Psi}{\partial x_a}+\frac{\partial \Psi}{\partial x_b} = 0.
\end{equation}
Defining $\hat{p}_a=-i\hbar\partial/\partial x_a$ and $\hat{p}_b=-i\hbar\partial/\partial x_b$, the above equation is rewritten as
\begin{equation}
\label{const4}
    (\hat{p}_a+\hat{p}_b) \Psi = 0.
\end{equation}
Eqs. \eqref{BPCSE} and \eqref{const4} are identical to \eqref{DiracSE} and \eqref{Qconstrained}, thus we obtain the same results as those using the Dirac quantization.

For the bipartite system with a global translational invariance constraint, we have shown that all three approaches, reduced quantization, Dirac quantization, and the framework based on the extended stationary action principle, yield the same result. However, this agreement is strongly dependent on the nature of the constraint. In this case, the constraint $\phi$ induced by the translation invariance is simple and can be recasted as a linear operator when it acts on the wave function.

\section{Discussion and conclusions} 
\label{sec:discussion}

\subsection{Limitation of the Dirac Constraints}
The limitation of the Dirac constraints \eqref{qconst} has been clearly explained in \cite{Hall}. In terms of position operator and momentum operator for a one-dimensional system, and substituting $\Psi=\sqrt{\rho}e^{iS/\hbar}$ into \eqref{qconst} since in our framework the dynamics of the system is expressed in terms of $S$ and $\rho$, we can re-expressed \eqref{qconst} as
\begin{equation}
\label{qconst2}
    \phi(x, -i\hbar\frac{\partial}{\partial x})\sqrt{\rho}e^{iS/\hbar} = 0.
\end{equation}
Thus, we should obtain two independent equations (the real and imaginary parts of \eqref{qconst2}).
\begin{equation}
\label{qconst3}
    R(S, \rho)=0; \text{ } I(S, \rho)=0.
\end{equation}
The above equations indicate that the Dirac constraint \eqref{qconst} is a strong condition that imposes two independent constraints for the conjugate pair $(S,\rho)$. However, the set of possible constraints on $(S,\rho)$ is much larger than those that can be expressed as \eqref{qconst2}, as the example shown in Section \ref{sec:ex1}. There, the constraints can be effectively expressed as
\begin{equation}
    \frac{\partial S}{\partial x}=0, \text{ } \frac{\partial \rho}{\partial t}=0.
\end{equation}
It is clear that they cannot be expressed in the linear form of \eqref{qconst2}. 

There is no justification that a physical constraint must always be expressed in the form of \eqref{qconst2}. This imposes a strong limitation on the Dirac quantization. On the other hand, the quantization framework presented in this paper is more generic and can naturally quantize those constraints that the Dirac quantization cannot handle. As mentioned in the Introduction section, although we arrive at the same conclusion as in \cite{Hall}, the mathematical framework for quantizing a constraint system is different. \cite{Hall} still adopts the approach of quantization first, and solving the constraint at the quantum level, where we propose here to quantize and constrain at the same time.

\subsection{On the Bohm Quantum Potential}
\label{sec:Bohm}
In Sec.~\ref{sec:ESAP}, we show that the time-independent Schrödinger equation~\eqref{TimeIndSE} can be recast as a rearrangement of the Bohm quantum potential $Q$. The dynamics governed by Eq.~\eqref{TimeIndSE} are therefore purely quantum in origin. Importantly, $Q$ arises directly from varying the information metric $I_f$ in Eq.~\eqref{FisherInfo} with respect to the probability density $\rho$. Hence, quantum effects emerge as the consequence of extremizing the relative entropy encoded in $I_f$. This provides an information-theoretic foundation for phenomena traditionally regarded as intrinsically quantum. For example, in Bohmian mechanics, $Q$ is postulated and attributed to hidden nonlocal variables, whereas in the present framework it is simply the variational outcome of $I_f$. The Bohmian “guiding wave” can thus be reinterpreted as the system’s tendency to minimize $I_f$ during its evolution. Moreover, for a bipartite system, the potentials $Q_a$ and $Q_b$ in Eq.~\eqref{BPQHJ} cannot, in general, be decomposed into independent contributions from the two subsystems. This inseparability originates from the structure of $I_f$, rather than from hidden variables, and offers a new perspective on the nature of entanglement, as further analyzed in Ref.~\cite{Yang2023_2}. %Furthermore, such an inseparability is preserved and propagated through the vacuum fluctuations of the bipartite system. Locality of the vacuum fluctuations rules out the need for a non-local mechanism. In summary, a mathematical term identical to the Bohm quantum potential is derived from the information metrics $I_f$, and our theory shows that the inseparability of this term is not associated with a non-local mechanism.

%\subsection{Limitations}
%Actual experiments to confirm Bell theorem~\cite{Hensen} are typically not based on entanglement of momentum or position, but based on other degrees of freedom such as electron spins, or polarization of photons. In order to explain the entanglement of such degrees of freedom from an information perspective, we need to define information metrics in addition to $I_f$ for these degrees of freedom, then apply the extended least action principle to derive quantum dynamics equations, for example, the Pauli equation for electron spins. This subject is beyond the current scope of this paper, and is reported in a separated paper~\cite{Yang2024}. Ref. \cite{Yang2024} shows that spin entanglement is the consequence of correlation between the random orientations of the intrinsic angular momenta of the two electrons. Since the orientation is an intrinsic local property of electron, the correlation of orientations can be preserved even when the two electrons are remotely separated. Such a correlation can be manifested without causal effect. The results further confirm the ideas presented in this paper.

\subsection{Conclusions}

In this work, we have developed a novel framework for the quantization of constrained systems, building on the extended stationary action principle. The key feature of this approach is that quantization and constraint enforcement are carried out simultaneously. This is achieved naturally within a variational framework, where constraints are incorporated directly into the Lagrangian through Lagrange multipliers. This approach removes the ambiguity of whether constraints should be imposed before or after quantization, a difficulty intrinsic to both reduced quantization and Dirac quantization.

To demonstrate the difference between the new quantization approach and the reduced quantization or Dirac quantization, we quantize two one-dimensional constrained systems with the three approaches. In the first example, we quantize a one-dimensional ensemble with vanishing local momentum. The reduced quantization and Dirac quantization give a trivial quantum state with constant probability distribution, while the new quantization approach yields a time-independent Schr\"{o}dinger equation for the constrained system, in addition to the trivial quantum state. For the bipartite system with global translational invariance, all three methods converge to the same quantum theory. These examples demonstrate that the present framework is consistent with the traditional approaches, while extending their applicability to constraints that cannot be represented as linear operators acting on the wave function.

The strategy of implementing quantization and constraining simultaneously is both mathematically natural and conceptually appealing. An important avenue for future work is to investigate extensions to more complex systems, including gauge theories and models of quantum gravity. Although such developments are beyond the scope of the present study, we believe that the variational framework introduced here provides a solid foundation for these future investigations.

\section*{Data Availability Statement}
Data supporting the findings of this study are available within the article.

%\textit{Acknowledgments.\,---\,} JMY thanks abc for helpful discussions. 

%\vspace*{1mm}
%\textit{Contributions.\,---\,} JMY designed the study, conceived the ideas, performed the mathematical calculation, and wrote the manuscript.

%========================================
%========================================

%========================================
%========================================

%\bibliography{RelFact} 

%========================================
%========================================
%========================================

\onecolumngrid

\pagebreak
%========================================

\appendix
\section{Extended Canonical Transformation}
\label{appendix:canonical}
In classical mechanics, the canonical transformation is a change of canonical coordinates $({x}, {p})$ to generalized canonical coordinates $({X}, {P})$ that preserves the form of Hamilton's equations. Without loss of generality,  here we restrict the formulation to one-dimensional spatial coordinate and momentum conjugate. Denote the Lagrangian for both canonical coordinate systems by $L_{xp}= {p}\dot{{x}}-H( {x}, {p},t)$ and $L'_{XP}={P}\dot{{X}}-K({X},{P},t)$, respectively, where $K$ is the new form of Hamiltonian with generalized coordinates. To ensure that the form of Hamilton's equations is preserved from the least action principle, one must have 
\begin{align}
    \delta \int^{t_B}_{t_A}dt L_{xp} &= \int^{t_B}_{t_A}dt ({p}\dot{{x}}-H({x},{p},t)) = 0,\\
    \delta \int^{t_B}_{t_A}dt L'_{XP} &= \int^{t_B}_{t_A}dt ({P}\dot{{X}}-K({X},{P},t)) = 0.
\end{align}
One way to meet such conditions is that the Lagrangians in both integrals satisfy the following relation
\begin{equation}
    \label{extCan}
    {P}\dot{{X}}-K({X},{P},t) = \lambda ({p}\dot{{x}}-H({x},{p},t)) + \frac{dG}{dt},
\end{equation}
where $G$ is a generating function and $\lambda$ is a constant. The justification here is that adding a total time derivative $dG/dt$ will not alter the dynamics equation of motion through the variation procedure over the action. There is a degree of freedom to choose the variables for the generating function $G$, which connects the new canonical coordinates to the old canonical coordinates. The variables for $G$ can be one of the old canonical coordinates $x$ or $p$, one of the new canonical coordinates $X$ or $P$, and possibly the time $t$. Thus, there are four types of possible canonical transformations~\cite{Goldstein}, depending on the choice of variables for $G$. In addition, $\lambda$ is another free parameter. When $\lambda \ne 1$, the transformation is called extended canonical transformations. Here we will choose $\lambda=-1$ so that the resulting Lagrangian takes the form that is related to the Hamilton-Jacobi equation, as will be seen shortly. Re-arranging (\ref{extCan}), we have
\begin{equation}
    \label{extCan2}
    \frac{dG}{dt} = {P}\dot{{X}} +{p}\dot{{x}} - (K+H).
\end{equation}
Choose a generating function $G={P}{{X}} + S({x}, {P}, t)$, a type 2 generating function~\cite{Goldstein}. Its total time derivative is
\begin{equation}
    \label{type1}
    \frac{dG}{dt} = {P}\dot{{X}} + {X}\dot{{P}} + \nabla S\dot{{x}} + \nabla_P S\dot{{P}} + \frac{\partial S}{\partial t}.
\end{equation}
The divergence operator $\nabla_P$ refers to a partial derivative over the generalized momenta ${P}$. Comparison of (\ref{extCan2}) and \ref{type1}) results in
\begin{align}
    \label{type12}
    \frac{\partial S}{\partial t} &= - (K+H), \\
    {p} &= \nabla S, \\
    {X} &= -\nabla_P S.
\end{align}
From (\ref{type12}), $K= - (\partial S/\partial t + H)$. Thus, $L'_{XP} = {P}\dot{{X}} + (\partial S/\partial t + H)$. We can choose a generating function $S$ such that ${X}$ does not explicitly depend on $t$ during motion. For instance, supposed $S({x}, {P}, t)=F({x}, {P}) + f({x}, t)$, one has ${X}=-\nabla_P F({x}, {P})$, so that $\dot{{X}}=0$ and $L'_{XP} = \partial S/\partial t + H({x}, {p}, t)$. Then the action integral in the generalized canonical coordinates becomes
\begin{equation}
    \label{extAction}
    A_c = \int^{t_B}_{t_A}dt L'_{XP} = \int^{t_B}_{t_A}dt \{\frac{\partial S}{\partial t} + H({x}, \nabla S, t)\}.
\end{equation}
In general, $P$ is not a constant, but can be inverted as a function of $x$ and $p$ through the equation ${p}= \nabla S$. However, if one further imposes a constraint on the generating function $S$ such that the generalized Hamiltonian $K=0$, Eq. (\ref{type12}) becomes the Hamilton-Jacobi equation $\partial S/\partial t + H = 0$, and both $P$ and $X$ are constants. The Hamilton-Jacobi equation is a special solution for the least action principle based on $A_c$ when $K=0$.

Note that choosing parameter $\lambda$ with a different value than $-1$ will not result in the formulation of $A_c$ in \eqref{extAction}. Eq. \eqref{extAction} expresses the classical action $A_c$ in such a form that when extending the formulation for the ensemble configurations allows us to derive the Hamilton-Jacobi equation. For an ensemble configuration with probability density $\rho({x}, t)$, the Lagrangian density $\mathcal{L}=\rho L'_{XP}$, and the average value of the classical action is,
\begin{equation}
    \label{extAction2}
    A_c = \int d{x}dt \mathcal{L} = \int d{x}dt \rho \{\frac{\partial S}{\partial t} + H({x}, \nabla S, t)\},
\end{equation}
Treating $(\rho, S)$ as the generalized canonical coordinates and momenta for the ensemble configuration, one can apply the least action principle with the above classical action $A_c$ to recover both the Hamilton-Jacobi equation and the continuity equation~\cite{Hall:2001,Hall:2002}.

\section{Consistency Checking of Constraints}
\label{appdx:constcheck}
Recall the definition of functional derivative for functional $F[f(x),g(x)]$
\begin{equation}
    \label{FD}
    \delta F[f,g]=\int dx\{(\frac{\delta F}{\delta f})\delta f+(\frac{\delta F}{\delta g})\delta g\}.
\end{equation}
Given \eqref{const24}, we have
\begin{equation}
    \delta\Phi_e = \int dx \{(\frac{\partial S}{\partial x} - p_c)\delta\rho  - \Delta\rho\delta S \}.
\end{equation}
Therefore, 
\begin{equation}
    \frac{\delta\Phi_e}{\delta\rho} = (\frac{\partial S}{\partial x} - p_c), \text{ }\frac{\delta\Phi_e}{\delta S}=- \Delta\rho.
\end{equation}
To check the consistency of the constraint \eqref{const24} with the equation of motion governed by the Hamiltonian \eqref{HC}, we evaluate the Poisson bracket on the constrained surface.
\begin{equation}
\begin{split}
    \{\Phi_e, H_e\} &= \int dx\{\frac{\delta\Phi_e}{\delta\rho}\frac{\delta H_e}{\delta S} - \frac{\delta\Phi_e}{\delta S}\frac{\delta H_e}{\delta \rho}\}\\
    &=\int dx \{(\frac{\partial S}{\partial x} - p_c)\frac{\delta H_e}{\delta S}+\frac{\partial\rho}{\partial x}(V+Q)\}\\
    & \approx -\int dx\rho \frac{\partial}{\partial x}(V+Q)\\
    & \approx -\int dx\rho \frac{\partial}{\partial x}(E)\approx 0.
\end{split}
\end{equation}
We utilize \eqref{StaticHJE} in step four of the above derivation. To verify $\{\Theta_e, H_e\}=0$, it is necessary to note that both $\Theta_e$ in \eqref{qconst24} and $H_e$ in \eqref{HC} are independent of $S$, so that $\delta\Theta_e/\delta S =0$ and $\delta H_e/\delta S =0$.

To check the consistency of the constraint $\Phi_e$ in \eqref{constrained3} with the equation of motion governed by the Hamiltonian $H_e$ in \eqref{HeBP}, we need to verify that the Poisson bracket $\{\Phi_e, H_e\}$ vanishes on the constrained surface. Given that
\begin{equation}
    \delta\Phi_e =\int dx_adx_b \{(\frac{\partial S}{\partial x_a}+\frac{\partial S}{\partial x_b} )\delta\rho - (\frac{\partial \rho}{\partial x_a}+\frac{\partial \rho}{\partial x_b} )\delta S\},
\end{equation}
we have
\begin{equation}
   \frac{ \delta\Phi_e}{\delta\rho} = \frac{\partial S}{\partial x_a}+\frac{\partial S}{\partial x_b}, \text{ }
   \frac{ \delta\Phi_e}{\delta S} =-(\frac{\partial \rho}{\partial x_a}+\frac{\partial \rho}{\partial x_b}).
\end{equation}
Similarly, one can calculate the following,
\begin{equation}
   \frac{ \delta H_e}{\delta\rho} = H, \text{ }
   \frac{ \delta H_e}{\delta S} =\frac{1}{m_a}\frac{\partial }{\partial x_a}(\rho\frac{\partial S}{\partial x_a}) + \frac{1}{m_b}\frac{\partial }{\partial x_b}(\rho\frac{\partial S}{\partial x_b}).
\end{equation}
Then, the Poisson bracket is
\begin{align*}
    \{\Phi_e, H_e\} & = \int dx_adx_b\{\frac{\delta\Phi_e}{\delta\rho}\frac{\delta H_e}{\delta S} - \frac{\delta\Phi_e}{\delta S}\frac{\delta H_e}{\delta \rho}\}\\
    &=\int dx_adx_b\{(\frac{\partial S}{\partial x_a}+\frac{\partial S}{\partial x_b})\frac{\delta H_e}{\delta S} + (\frac{\partial \rho}{\partial x_a}+\frac{\partial \rho}{\partial x_b})H\} \\
    &\approx -\int dx_adx_b\{\rho(\frac{\partial H}{\partial x_a}+\frac{\partial H}{\partial x_b})\} \approx 0.
\end{align*}
We use \eqref{constrained2} recursively in the last two steps of the derivation above. Thus, the constraint \eqref{constrained2} does not induce secondary constraints. 

\section{Quantization of a Bipartite System}
\label{appx:bipartite}
We follow the five steps described in Section \ref{LIP} to quantize a bipartite one-dimensional system consisting two subsystems $a$ and $b$. For such a system, the Lagrangian is
\begin{equation}
    L = \frac{1}{2}mv_a^2 +  \frac{1}{2}mv_b^2 - V(x_a, x_b).
\end{equation}
In Step II, we seek to derive the probability distribution of bipartite system for the vacuum fluctuations from positions $(x_a, x_b)$ to $(x_a+w_a, x_b+w_b)$ in an infinitesimal time interval $\Delta t$. This is achieved by defining
\begin{align*}
    I_f  =: D_{KL}(\varrho({x}_a+{w}_a,{x}_b+{w}_b |{x}_b) || \sigma) 
    = \int \varrho \ln(\varrho/\sigma)dw_adw_b.
\end{align*}
where $\sigma$ is a uniform probability distribution. The classical action of the ensemble in the time interval $\Delta t$ is
\begin{equation}
    A_c=\int\varrho\{\frac{m_a}{2\Delta t}w_a^2 + \frac{m_b}{2\Delta t}w_b^2+ V(x_a, x_b)\Delta t\} dw_adw_b.
\end{equation}
The third term can be ignored when $\Delta t\to 0$. Thus, the total action is
\begin{equation}
\label{A11}
    A_t=\int\varrho\{\frac{m_a}{2\Delta t}w_a^2 + \frac{m_b}{2\Delta t}w_b^2 +\frac{\hbar}{2}\ln\frac{\varrho}{\sigma}\} dw_adw_b.
\end{equation}
Performing variation of $A_c$ with respect to $\varrho$, one obtains
\begin{equation}
    \varrho(w_a, w_b) = \frac{1}{Z}\exp\{-\frac{m_a}{\hbar\Delta t}w_a^2-\frac{m_b}{\hbar\Delta t}w_b^2\},
\end{equation}
where $Z$ is a normalization factor. This transition probability distribution can be rewritten in a separated form,
\begin{equation}
\label{BPp}
    \varrho(w_a, w_b) = \varrho_a(w_a)\varrho_b(w_b)=(\frac{1}{Z_a}\exp\{-\frac{m_a}{\hbar\Delta t}w_a^2\})(\frac{1}{Z_b}\exp\{-\frac{m_b}{\hbar\Delta t}w_b^2\}).
\end{equation}
This is consistent with Assumption 1 that the vacuum fluctuation is completely local. From \eqref{BPp}, one can verify that
\begin{equation}
    \label{BPvariance>}
    \langle w_a\rangle=\langle w_a\rangle=0,\textbf{ }\langle w_a^2\rangle = \frac{\hbar\Delta t}{2m_a}, \textbf{ } \langle w_b^2\rangle = \frac{\hbar\Delta t}{2m_b}.
\end{equation}

To perform the classical canonical transformation in Step III, we introduce two pairs of generalized canonical coordinates $\{X_a, P_a\}$ and $\{X_b, P_a\}$. Denote the Hamiltonian in the generalized canonical coordinate system as $K(X_a, P_a, X_b, P_B)$. The canonical transformation requires
\begin{equation}
\label{BPCanT}
    P_a\dot{X}_a+P_b\dot{X}_b-K(X_a,P_a,X_b,P_b)
    = \lambda (p_a\dot{x}_a+p_b\dot{x}_b-H(x_a,p_a,x_b,p_b)) + \frac{dG}{dt},
\end{equation}
where $G$ is a generating function. Choosing $\lambda=-1$ and a type 2 generating function $G=P_aX_a+P_bX_b+S(x_a, x_b, P_a, P_b, t)$, computing the total time derivative $dG/dt$ and comparing it with \ref{BPCanT}, one finds that
\begin{align}
    \label{BPT2}
    \frac{\partial S}{\partial t} &= - (K+H), \\
    p_a &= \frac{\partial S}{\partial x_a}, \text{ } p_b= \frac{\partial S}{\partial x_b},\\
    X_a &= -\frac{\partial S}{\partial P_a}, \text{ } X_b=-\frac{\partial S}{\partial P_b}.
\end{align}
Choosing the generating function $S$ such that $\dot{X}_a=0$ and $\dot{X}_b=0$, we have the Lagrangian in the generalized canonical coordinate system as $L'=P_a\dot{X}_a+P_b\dot{X}_b-K =(\partial S/\partial t +H)$. Thus, the classical action for the bipartite system is
\begin{equation}
    \label{BPAction}
    A_c = \int dt \{\frac{\partial S}{\partial t} + \frac{1}{2m_a}(\frac{\partial S}{\partial x_a})^2 + \frac{1}{2m_b}(\frac{\partial S}{\partial x_b})^2 + V(x_a, x_b)\}.
\end{equation}
For the ensemble system with probability density $\rho(x_a, x_b, t)$, the Lagrangian density $\mathcal{L}=\rho L'$, and the average value of the classical action of the bipartite ensemble is
\begin{equation}
    \label{extActionB}
    A_c =\int dx_adx_bdt \rho(x_a,x_b,t) \{\frac{\partial S}{\partial t} + \frac{1}{2m_a}(\frac{\partial S}{\partial x_a})^2 + \frac{1}{2m_b}(\frac{\partial S}{\partial x_b})^2 + V(x_a, x_b)\}.
\end{equation}
It is important to note that $\rho(x_a,x_b,t)\ne \rho_a(x_a,t)\rho_b(x_b,t)$. That is, the probability density for the bipartite ensemble is not necessarily separable. This fact is crucial to understanding the potential entanglement between the two subsystems $A$ and $B$. Fixed point variation of $A_c$ with respect to $\rho$ gives the Hamilton-Jacobi equation,
\begin{equation}
    \frac{\partial S}{\partial t} + \frac{1}{2m_a}(\frac{\partial S}{\partial x_a})^2 + \frac{1}{2m_b}(\frac{\partial S}{\partial x_b})^2 + V(x_a, x_b) = 0,
\end{equation}
and variation with respect to $S$ gives the continuity equation,
\begin{equation}
\label{BPCont}
    \frac{\partial \rho}{\partial t} + \frac{1}{m_a}\frac{\partial}{\partial x_a}(\rho\frac{\partial S}{\partial x_a})^2 + \frac{1}{m_b}\frac{\partial}{\partial x_b}(\rho\frac{\partial S}{\partial x_b})^2 = 0.
\end{equation}

In Step IV, we extend the definition of $I_f$ in (\ref{DLDivergence}) to the bipartite system:
\begin{align*}
\label{DLDivergencefor2}
    I_f &=: \sum_{j=0}^{N-1}\langle D_{KL}(\rho ({x}_a,{x}_b, t_j) || \rho ({x}_a+{w}_a, {x}_b+{w}_b, t_j)\rangle_{w_a,w_b} \\
    &=\sum_{j=0}^{N-1}\int d{w}_ad{w}_b \varrho_a({w}_a)\varrho_b({w}_b)\int d{x}_ad{x}_b\rho ({x}_a,{x}_b, t_j)\ln \frac{\rho ({x}_a,{x}_b, t_j)}{\rho ({x}_a+{w}_a, {x}_b+{w}_b, t_j)}.
\end{align*}
Expanding the logarithmic function up to $O(w_a^2, w_b^2)$,
\begin{align}
    \ln\frac{\rho ({x}_a,{x}_b, t_j)}{\rho ({x}_a+{w}_a, {x}_b+{w}_b, t_j) } = \frac{1}{\rho}(-\frac{\partial\rho}{\partial x_a}w_a-\frac{1}{2}\frac{\partial^2\rho}{\partial x_a^2}w_a^2+\frac{1}{2\rho}(\frac{\partial\rho}{\partial x_a})^2w_a^2-\frac{\partial\rho}{\partial x_b}w_b-\frac{1}{2}\frac{\partial^2\rho}{\partial x_b^2}w_b^2+\frac{1}{2\rho}(\frac{\partial\rho}{\partial x_b})^2w_b^2),
\end{align}
we have
\begin{align*}
    I_f =& \sum_{j=0}^{N-1}\int d{x}_ad{x}_b(-\frac{\partial\rho}{\partial x_a}\langle w_a\rangle -\frac{1}{2}\frac{\partial^2\rho}{\partial x_a^2}\langle w_a^2\rangle + \frac{1}{2\rho}(\frac{\partial\rho}{\partial x_a})^2\langle w_a^2\rangle-\frac{\partial\rho}{\partial x_b}\langle w_b\rangle -\frac{1}{2}\frac{\partial^2\rho}{\partial x_b^2}\langle w_b^2\rangle + \frac{1}{2\rho}(\frac{\partial\rho}{\partial x_b})^2\langle w_b^2\rangle)\\
    =&\sum_{j=0}^{N-1}\int d{x}_ad{x}_b(-\frac{1}{2}\frac{\partial^2\rho}{\partial x_a^2}\langle w_a^2\rangle + \frac{1}{2\rho}(\frac{\partial\rho}{\partial x_a})^2\langle w_a^2\rangle -\frac{1}{2}\frac{\partial^2\rho}{\partial x_b^2}\langle w_b^2\rangle + \frac{1}{2\rho}(\frac{\partial\rho}{\partial x_b})^2\langle w_b^2\rangle)
\end{align*}
The last step uses the fact that $\langle w_{ia}\rangle = \langle w_{ib}\rangle = 0$. Taking the assumption that $\rho$ is a regular function and its gradient with respect to ${x}_a$ or ${x}_b$ approaches zero when $|{x}_a|, |{x}_b|\to\pm\infty$, the first and third terms vanish.
Substituting $\langle w_{a}^2\rangle = \hbar\Delta t/2m_a$ and $\langle w_{b}^2\rangle = \hbar\Delta t/2m_b$, and taking $\Delta t\to 0$, we get
\begin{equation}
\label{If2}
    I_f = \int d{x}_ad{x}_bdt\{\frac{\hbar}{4m_a}\frac{1}{\rho}(\frac{\partial\rho}{\partial x_a})^2 + \frac{\hbar}{4m_b}\frac{1}{\rho}(\frac{\partial\rho}{\partial x_b})^2\}.
\end{equation}
Now we have the total action of the bipartite ensemble
\begin{equation}
\label{A17}
    A_t = \int dx_adx_bdt\rho \{\frac{\partial S}{\partial t} + \frac{1}{2m_a}(\frac{\partial S}{\partial x_a})^2 + \frac{1}{2m_b}(\frac{\partial S}{\partial x_b})^2 + V+\frac{\hbar}{4m_a}(\frac{\partial \ln\rho}{\partial x_a})^2 + \frac{\hbar}{4m_b}(\frac{\partial\ln\rho}{\partial x_b})^2\}.
\end{equation}
Step V, taking variation of $A_t$ with respect to $S$ leads to the same continuity equation as \eqref{BPCont}, while performing variation with respect to $\rho$ gives the quantum version of Hamilton-Jacobi equation
\begin{equation}
\label{BPQHJ}
    \frac{\partial S}{\partial t} + \frac{1}{2m_a}(\frac{\partial S}{\partial x_a})^2 +\frac{1}{2m_b}(\frac{\partial S}{\partial x_v})^2+ V +Q_a+Q_b=0, \text{ where } Q_\alpha =- \frac{\hbar^2}{2m_\alpha}\frac{\nabla_\alpha^2\sqrt{\rho}}{\sqrt{\rho}} \text{ and }\{\alpha=a,b\}.
\end{equation}
Defining a complex function $\Psi(x_a,x_b,t)=\sqrt{\rho(x_a,x_b,t)}e^{iS(x_a,x_b,t)/\hbar}$, the continuity equation and the extended Hamilton-Jacobi equation (\ref{BPQHJ}) can be combined into a single differential equation, the Schr\"{o}dinger equation,
\begin{equation}
    i\hbar\frac{\partial\Psi}{\partial t} = [-\frac{\hbar^2}{2m_a}\nabla_a^2 -\frac{\hbar^2}{2m_b}\nabla_b^2+ V]\Psi.
\end{equation}

%Equation (\ref{If2}) shows that $I_f$ is inseparable since $\rho({x}_a, {x}_b, t) \ne  \rho_a({x}_a, t)\rho_b({x}_b, t)$. On the other hand, suppose $\rho({x}_a, {x}_b, t) =  \rho_a({x}_a, t)\rho_b({x}_b, t)$, then $\nabla_a \rho = \rho_b\nabla_a\rho_a$. Similarly, $\nabla_b \rho = \rho_a\nabla_b\rho_b$, then
%\begin{align}
%    I_f &= \int d{x}_ad{x}_bdt\{\frac{\hbar}{4m_a}\frac{\nabla_a\rho_a\cdot\nabla_a\rho_a}{\rho_a}\rho_b + \frac{\hbar}{4m_b}\frac{\nabla_b\rho_b\cdot\nabla_b\rho_b}{\rho_b}\rho_a\} \\
%    &= \frac{\hbar}{4m_a}\int d{x}_adt\frac{\nabla_a\rho_a\cdot\nabla_a\rho_a}{\rho_a} + \frac{\hbar}{4m_b}\int d{x}_bdt\frac{\nabla_b\rho_b\cdot\nabla_b\rho_b}{\rho_b} = (I_f)_a + (I_f)_b,
%\end{align}
%and is clearly separable into two independent terms, where 
%\begin{equation}
%    (I_f)_a=\frac{\hbar}{4m_a}\int d{x}_adt\frac{\nabla_a\rho_a\cdot\nabla_a\rho_a}{\rho_a}; \text{   } (I_f)_b = \frac{\hbar}{4m_b}\int d{x}_bdt\frac{\nabla_b\rho_b\cdot\nabla_b\rho_b}{\rho_b}.
%\end{equation}

\end{document}